# Optical response of a self-standing monolayer of dielectric spheres

Luis A. Dorado and Ricardo A. Depine*

Grupo de Electromagnetismo Aplicado, Departamento de Física, Facultad de Ciencias Exactas y Naturales, Universidad de Buenos Aires, Buenos Aires, Argentina. *rdep@df.uba.ar

**Abstract**: An analysis of the optical response of periodically arrayed monolayers composed of dielectric spheres with low refractive index is herein presented. The reflectance spectra of two-dimensional square and triangular lattices are obtained by means of the vector Korringa-Kohn-Rostoker method, both spectra showing very similar qualitative features for photon energies below the onset of diffraction spots. In this energy region, the same number of peaks of unitary amplitude in the reflectance spectra are predicted for both kinds of monolayers, suggesting that this must be a universal feature independent of the particular geometry of the lattice. The origin of these high reflectance peaks is investigated. It is found that the resonances of TM and TE modes due to dipolar, quadrupolar and octupolar interaction inside the monolayer are largely responsible for the peak structure observed in the reflectance spectra.

The analysis of the optical response of a monolayer composed of a dielectric material with two-dimensional (2D) periodicity is paramount in understanding the behaviour of a more complex structure formed by piling monolayers up in order to build a three-dimensional (3D) periodicity.[1,2,3] Slabs of colloidal photonic crystals formed by dielectric spheres have attracted much attention in the last decades, mainly due to the advent of easier processes of fabrication, especially in the micrometer scale. Subsequent improvements in these techniques have also allowed the achievement of high quality crystal slabs with interesting optical features.[4,5,6] Despite the high complexity of the scattering phenomena that take place inside a monolayer,[7,8,9] the optical response of the system can be understood by analysing two limiting cases, namely, those of spheres with either high or low refractive index.

When the refractive index of the spheres is high, a tight-binding approximation can be used since photons tend to concentrate inside the spheres, whose internal field distribution is very similar to that of a single isolated sphere.[10] Hence, this approximation allows us to recognize that the single-sphere Mie resonances[11] are responsible for the dips in the transmittance spectrum (peaks in the reflectance spectrum) of the monolayer in this case. On the contrary, if the refractive index of the spheres is low, the single-sphere Mie resonances are far from those observed in the optical spectra since photons are almost free. Under this condition, the band structure can be obtained by means of a nearly-free-photon approximation, which is based on the folding of the free-photon dispersion relation into the first Brillouin zone.[12] In this case, effective-medium theory could be applied by replacing the sphere monolayer with an effective homogeneous medium in order to predict the reflectance and transmittance spectra. This simple method provides a satisfactory explanation of the spectra features at low energies. Nevertheless, this limiting case of low refractive index can no longer be applied at wavelengths comparable to the diameter of the spheres. Furthermore, the internal field distribution of the monolayer with low dielectric constant is complicated and strongly influenced by the interaction between spheres.[10]

In this work, we have focused our interest on the optical response of a monolayer composed of spheres with low dielectric constant for those energies near the onset of diffraction spots, where the tight-binding and nearly-free-photon approximations break down and cannot be used in order to explain the electromagnetic response of the system satisfactorily. However, these approaches have provided a remarkable physical insight within their respective ranges of validity.[13,14] The approach we will follow takes into account the 2D periodicity of the monolayer and its internal electrodynamics, which can be understood in terms of multipolar interactions between the dielectric spheres centred at sites of a 2D Bravais lattice. Therefore, resonance phenomena inside the monolayer will be addressed by approximating the spheres in the lattice by electric and magnetic dipoles, quadrupoles and octupoles, which are the lowest order multipoles determining the optical spectra features of the monolayer with low dielectric constant.

When we consider an electromagnetic wave incident on a sphere of radius S, dielectric constant $\varepsilon_s$ and magnetic permeability $\mu_s$, centred at the origin of coordinates ($\mathbf{r} = 0$), the spherical wave expansions of the incident electric and magnetic fields around the origin are given by[15,16]

$$\mathbf{E}(\mathbf{r}) = \sum_{l=1}^{\infty} \sum_{m=-l}^{l} \left( \frac{i}{k} a_{lm}^{0E} \nabla \times j_l(kr) \mathbf{X}_{lm}(\theta, \phi) + a_{lm}^{0H} j_l(kr) \mathbf{X}_{lm}(\theta, \phi) \right), \quad (1)$$

$$\mathbf{H}(\mathbf{r}) = \sqrt{\frac{\varepsilon \varepsilon_0}{\mu \mu_0}} \sum_{l=1}^{\infty} \sum_{m=-l}^{l} \left( a_{lm}^{0E} j_l(kr) \mathbf{X}_{lm}(\theta, \phi) - \frac{i}{k} a_{lm}^{0H} \nabla \times j_l(kr) \mathbf{X}_{lm}(\theta, \phi) \right), \quad (2)$$

where $(r, \theta, \phi)$ are the usual spherical coordinates of an evaluation point $\mathbf{r}$, $\varepsilon$ and $\mu$ are the dielectric constant and magnetic permeability of the homogeneous medium outside the sphere, $k = \sqrt{\mu\varepsilon}\,\omega/c$ is the magnitude of the wavevector, $c = 1/\sqrt{\mu_0 \varepsilon_0}$ is the velocity of light in vacuum, $j_l(kr)$ are spherical Bessel functions of the first kind and $\mathbf{X}_{lm}(\theta, \phi)$ are vector spherical harmonics.

In the spherical wave expansion of the electromagnetic field, given by equations (1) and (2), each pair of integers $(l, m)$, where $l \geq 1$ and $-l \leq m \leq l$, represents a spherical partial wave of order $(l, m)$, which can be either electric (with complex amplitude $a_{lm}^{0E}$) or magnetic (with complex amplitude $a_{lm}^{0H}$). Dipolar fields correspond to $l = 1$ and are obtained by summing up three electric and magnetic partial waves in the series expansion, quadrupolar fields correspond to $l = 2$ and are obtained as the sum of five electric and magnetic partial waves, and so on. In general, a $2^l$-pole has $2l+1$ electric and magnetic terms in the multipole expansion. By using the orthogonality properties of the vector spherical harmonics,[15] it can easily be proved that electric multipoles have transverse magnetic (TM) fields, while magnetic multipoles have transverse electric (TE) fields. Therefore, this multipole expansion can also be viewed as a linear combination of TM and TE spherical waves.

The electromagnetic field scattered by the sphere at the origin can also be expanded in spherical waves and it is given by replacing $j_l(kr)$ with the spherical Hankel function of the first kind, $h_l^+(kr)$, in equations (1) and (2), which corresponds to an outgoing spherical wave from the origin. Also, the coefficients $a_{lm}^{0E}$ and $a_{lm}^{0H}$ must be replaced with the new coefficients $a_{lm}^{+E}$ and $a_{lm}^{+H}$ of the scattered wave. The fields inside the sphere have another set of coefficients, $a_{lm}^{IE}$ and $a_{lm}^{IH}$, and are obtained by replacing $k$ with the amplitude of the wavevector inside the sphere, $k_s = \sqrt{\mu_s \varepsilon_s}\, \omega/c$, in equations (1) and (2). Under these conditions, we can find the relationships between the coefficients $a_{lm}^{0E(H)}$ of the incident wave and the coefficients $a_{lm}^{+E(H)}$ of the scattered wave by forcing the continuity of the tangential components of the electric and magnetic fields at the surface of the sphere. Then, it follows that[17]

$$a_{lm}^{+E(H)} = T_l^{E(H)} a_{lm}^{0E(H)}, \qquad (3)$$

where

$$T_l^E = \left[\frac{j_l(k_s r)\frac{\partial}{\partial r}(r j_l(kr))\varepsilon_s - j_l(kr)\frac{\partial}{\partial r}(r j_l(k_s r))\varepsilon}{h_l^+(kr)\frac{\partial}{\partial r}(r j_l(k_s r))\varepsilon - j_l(k_s r)\frac{\partial}{\partial r}(r h_l^+(kr))\varepsilon_s}\right]_{r=S} , \quad (4)$$

$$T_l^H = \left[\frac{j_l(k_s r)\frac{\partial}{\partial r}(r j_l(kr))\mu_s - j_l(kr)\frac{\partial}{\partial r}(r j_l(k_s r))\mu}{h_l^+(kr)\frac{\partial}{\partial r}(r j_l(k_s r))\mu - j_l(k_s r)\frac{\partial}{\partial r}(r h_l^+(kr))\mu_s}\right]_{r=S} . \quad (5)$$

We can see that $T_l^E$ describes the scattered TM field of an electric $2^l$-pole and $T_l^H$ describes the scattered TE field of a magnetic $2^l$-pole. The total scattered field is obtained by summing the TM and TE contributions of all $2^l$-poles, from $l = 1$ to infinity. In a numerical calculation, a cut-off $L_{MAX}$ in this spherical wave expansion is chosen, retaining the first $L_{MAX}(L_{MAX} + 2)$ TM and TE terms. Also, in the calculation of the multiple scattering between spheres in a given 2D lattice, expressions for $a_{lm}^{IE}$ and $a_{lm}^{IH}$ are not required. In this case, the total incident wave on the sphere at the origin is given by the sum of an external incident plane wave and the waves scattered by all the other spheres in the lattice. This is the spirit of the vector Korringa-Kohn-Rostoker (KKR) method, which can be used for the calculation of the optical spectra and band structure of colloidal crystals and, as we can see, it is based on an accurate formalism where a rigorous solution to Maxwell's equations is obtained.[16,18]

In the layer version of the vector KKR method,[17,19] the 3D crystal slab is first divided into layers parallel to a given crystallographic plane, each 2D layer containing periodically arrayed identical spheres. Next, the multipole expansion in spherical waves is used to account for the multiple scattering process inside a layer. Finally, a series expansion in a plane wave basis permits us to describe the electromagnetic interactions between layers and to compute the far-field power flux in either side of the crystal slab.

In what follows, we will present an analysis of the behaviour of the optical spectra features of dielectric spheres arrayed in square and triangular 2D lattices by applying the vector KKR method. The lattice constant is *a* (nearest neighbours distance between sphere centres) and energy is expressed in reduced units $a/\lambda$ (normalized frequency), where $\lambda$ is the wavelength of the incident light. The sphere radius and dielectric constant are *S* and $\varepsilon_s$, respectively. The spheres are embedded in air ($\varepsilon$ = 1), which is also the medium in both sides of the monolayer. The onset of diffraction spots from this self-standing monolayer depends on the 2D periodicity of the lattice and the angle of incidence of the incoming light beam. However, we will focus on the optical properties for normal incidence, that is, when light impinges in the normal direction to the plane of the lattice.

In the reciprocal space of a given 2D lattice, any reciprocal lattice vector **g** is determined by a pair of integers (*p,q*) and, in the case of normal incidence, the diffraction cut-off can be obtained from $|\mathbf{g}| = 2\pi/\lambda$. By calculating the reciprocal lattice vectors of the 2D lattice,[20] it is straightforward to prove that the nearest neighbours distance in reciprocal space for a square lattice is given by $|\mathbf{g}| = 2\pi/a$, while we have $|\mathbf{g}| = 4\pi/(\sqrt{3}\,a)$ for a triangular lattice. Then, we can predict that the first diffraction spots appear at $a/\lambda$ = 1 for a square lattice (four diffracted beams) and at $a/\lambda = 2/\sqrt{3} \cong$ 1.155 for a triangular lattice (six diffracted beams). The ratio between the power carried by each diffracted mode and the incident power is given by a reflection coefficient $R_{p,q}$, where the same (*p,q*) used for **g** indicates the order of diffraction, so that (*p,q*)=(0,0) corresponds to the specularly reflected light, a quantity that is usually measured. The total reflectance, R, is the sum of the intensities of all diffracted modes in the homogeneous incoming medium plus the specularly reflected beam, that is, $R = \Sigma_{(p,q)} R_{p,q}$. For energies below the first diffraction cut-off, the total and specular reflectance coefficients are equal, $R = R_{0,0}$, because the specular beam (**g** = 0) is the only one that survives in the far-field zone (its wavevector is the only one with a real component normal to the surface of the monolayer).

Figure 1 shows the specular and total reflectance spectra, $R_{0,0}$ and R, of a square lattice of close-packed spheres of refractive index $n_s$ = 1.58 (dielectric constant $\varepsilon_s$ = 2.4964) in air, which would correspond to latex[4,5] ($n_s \cong$ 1.59). Also, the total reflected diffracted light spectrum, given by $\Sigma_{(p,q)\neq(0,0)} R_{p,q}$, can be seen in Fig. 1(b), where the onset of diffraction at $a/\lambda$ = 1 is indicated. The reduction of the specular reflectance for energies above the diffraction cut-off can clearly be seen, since four diffracted spots remove part of the energy from the specularly reflected beam. A similar behaviour can be appreciated in Fig. 2 for a triangular lattice of the same close-packed spheres used in Fig. 1. The total reflected diffracted light spectrum is plotted in Fig. 2(b), where the onset of diffraction at $a/\lambda \cong$ 1.155 is also indicated. In this case, six diffracted beams produce the intensity reduction of the specular reflectance, $R_{0,0}$.

In Figs. 1 and 2, four alternating broad and sharp reflectance peaks of unit amplitude can be observed below the energy of the diffraction cut-off in each lattice. In the case of the triangular lattice, the first two maxima in the reflectance spectrum, located at about $a/\lambda$ = 0.83, can be observed in the inset of Fig. 2(a), although they are difficult to appreciate at first sight. These similar qualitative features between both kinds of lattices indicate that there must be a universal underlying behaviour independent of the particular geometry of the lattice. As a matter of fact, by varying the angle between the primitive lattice vectors, we have found a similar pattern of reflectance peaks for all the lattice geometries obtained in this way.

For spheres of high refractive index arrayed in a triangular lattice, Kurokawa et al.[10] have shown that broad peaks originate from the Mie resonance of TM modes of an isolated sphere, while sharp peaks are near the single-sphere Mie resonance of TE modes. However, as we have already mentioned above, the Mie resonances of a single isolated sphere of low refractive index are far from the peaks observed in the reflectance spectrum (dips in transmittance spectrum) and, therefore, this tight-binding approximation cannot explain the origin of these reflectance peaks. Another extreme case is the nearly-free-photon approximation, which applies in the case of small spheres of low refractive index. In this case, effective-medium theory is sufficient to describe the optical response of the system at low energies. Nevertheless, as the photon energy approaches that of the onset of diffraction, the nearly-free-photon approach can no longer be used, since it does not predict the alternating pattern of peaks observed in the reflectance spectrum. Consequently, the peak structure observed in the spectra must be analysed by taking into account the multiple scattering between the spheres in the lattice and not just the Mie resonance of a single isolated sphere of low refractive index. Despite the high complexity of the electrodynamics inside the monolayer, we will show that the origin of the peak structure in the optical spectra can be understood in terms of resonances of TM and TE modes of the lattice structure considered as a whole entity.

In Fig. 3, the shift of the reflectance spectrum towards higher energies can be seen for a square lattice as the sphere radius is gradually decreased from $S = 0.5a$ (close-packed monolayer) to $S = 0.15a$. Figure 3(a) shows that only the first two peaks remain for energies below the onset of diffraction when the sphere radius decreases. The zoom shown in Fig. 3(b) for the small range $0.98 \leq a/\lambda \leq 1.0$ confirms that only two peaks are present when the sphere radius is small. The same behaviour of the reflectance spectra for decreasing values of the sphere radius can be observed in Fig. 4 for the triangular lattice. Very similar qualitative features can be appreciated for both lattices by comparing the spectra of Figs. 3(b) and 4(b) for small spheres. We can observe that the square lattice has the peaks closer to the onset of diffraction than in the case of the triangular lattice if a relative distance in energy is considered. For example, for spheres of radius $S = 0.2a$, the first peak of the square lattice is located at $a/\lambda = 0.987$ and the diffraction begins at $a/\lambda = 1.0$, then the relative position of the first peak with respect to the diffraction cut-off is given by $(1.0-0.987)/1.0 = 0.013$. In the case of the triangular lattice, the first peak is at $a/\lambda = 1.118$ while the onset of diffraction is at $a/\lambda$ = 1.155, then the relative position of this peak is (1.155-1.118)/1.155 = 0.032, which is more than twice as great than the relative distance for the square lattice. Because this effect is the same for any sphere radius, as shown in Figs. 3(a) and 4(a) for larger spheres, it must be somehow related to the geometry of the lattice. The square lattice has four nearest neighbours, while the triangular lattice has six of them, then electromagnetic interactions between spheres must be stronger for the latter than for the former. Despite the fact that the lead role of geometry in determining the resonance frequency is well known for simpler systems, e.g. the single-sphere Mie resonances, further theoretical work is necessary to clearly understand the role that the geometry of the lattice plays in the frequency shifts of the optical spectra features of the monolayers.

In order to reveal the origin of the reflectance peaks in Figs. 3 and 4, Figs. 5(a) and 5(c) show the reflectance spectra for the square and triangular lattices calculated with a cut-off $L_{MAX}$ = 9 in the spherical wave expansion, which is sufficient to obtain good convergence in the range of energies we are interested in. The reflectance spectrum obtained using $L_{MAX}$ = 1, which corresponds to dipolar interaction, is also plotted in Figs. 5(a) and 5(c). By considering $L_{MAX}$ = 1, the spheres in the lattice are replaced by electric and magnetic dipoles, then Fig. 5 shows that the monolayer is well approximated by a lattice of interacting dipoles when the sphere size is sufficiently small ($S$ = 0.2$a$, the smaller the spheres, the more accurate the dipolar approximation). In this case, only the coefficients $T_1^E$ and $T_1^H$ are needed to describe the scattered wave of each sphere in the lattice, which is a linear superposition of TM and TE modes. If we arbitrarily set $T_1^H = 0$, TE modes are automatically eliminated, that is, only electric dipoles, having TM fields, are considered. The reflectance spectra calculated using this TM dipolar contribution is shown by the black solid line in Fig. 5(b) for the square lattice and in Fig. 5(d) for the triangular lattice. We can immediately see that the first broad peak present in the reflectance spectrum of both lattices in Figs. 5(a) and 5(c) is a resonance of TM modes, which means that electric dipolar interaction inside the monolayer is responsible for this broad peak. Obviously, the vector sum of the TM fields produced by the electric dipoles in the lattice is a TM mode of the monolayer as a whole.

If we now arbitrarily set $T_1^E = 0$, all TM modes are eliminated and only magnetic dipoles, which have TE fields, are retained. In this case, the red dashed line in Figs. 5(b) and 5(d) shows the contribution of these TE modes to the reflectance spectrum of the square and triangular lattices, respectively. This means that if each sphere in the lattice is replaced by a magnetic dipole, a resonance phenomenon occurs, which manifests itself as a sharp peak in the reflectance spectra shown in Fig. 5. We must point out that when higher order multipoles are added to the description of the system, none other resonances appear because for values of $L_{MAX}$ greater than one the number of reflectance peaks remains unchanged, so dipoles produce only two reflectance peaks, with no new resonances when more multipoles are considered. We should also point out that the total reflectance R shown in Figs. 5(a) and 5(c) is not the sum of the TM and TE contributions shown in Figs. 5(b) and 5(d), because electric-magnetic dipole interactions have been neglected in the latter. Nevertheless, these results demonstrate that the TM and TE modes are almost uncoupled in this case.

At this point in our discussion, if we increase the sphere radius, the resonance peaks will naturally be shifted towards lower energies. Also, as the sphere size is gradually increased, electric and magnetic quadrupoles must be added in order to reproduce the real shape and position of the reflectance peaks, but dipolar interaction is still the main cause of these resonances. This effect can be seen in Fig. 6 for close-packed square and triangular lattices (*S* = 0.5*a*). We can observe that there are no new resonances related to quadrupolar interaction because when quadrupoles are added to the description of the system ($L_{MAX}$ = 2), no other reflectance peaks appear. However, quadrupole terms are essential to reproduce the real shape and position of these reflectance peaks.

In Fig. 6, two additional reflectance peaks near the diffraction cut-off can be observed for the close-packed lattices calculated with $L_{MAX}$ = 9, which is practically the real response of the system. These spectra are the same as the ones previously discussed and shown in Figs. 1 and 2 in a wider range of energies. As we have found in the spectra of Fig. 5 for small spheres, quadrupoles do not produce any of these two new peaks, as the blue dotted curve for $L_{MAX}$ = 2 demonstrates in Figs. 6(a) and 6(c) for the square and triangular lattices. Nevertheless, when octupoles are added ($L_{MAX}$ = 3), two new resonances appear very near to the two reflectance peaks we are discussing. Therefore, these peaks are resonances of TM and TE modes related to octupolar interaction inside the monolayer. As before, if we arbitrarily set $T_3^H = 0$, magnetic octupoles with TE fields are eliminated and the remaining TM modes associated to electric octupoles, together with the existing dipoles and quadrupoles, are the ones that originate the second broad peak observed in the reflectance spectra. Conversely, by setting $T_3^E = 0$, only magnetic octupoles are retained and the second sharp peak is a resonance of their TE modes.

From this analysis, it follows that at each reflectance peak the system is in an eigenstate that is practically a linear superposition of the first three TM and TE multipole fields, which are dominant in monolayers with spheres of low refractive index, so broad peaks can be characterized as TM in origin while sharp peaks can be characterized as TE in origin, according to the multipole interaction responsible for the appearance of these peaks. Exactly the same happens for monolayers of spheres with high refractive index, where the reflectance peaks are either TM or TE in origin according to the closest Mie resonance mode of a single isolated sphere. Thus, the eigenstate of the system in resonance is not purely TM nor TE but a linear combination of them, in which one or the other predominates.

For the reasons stated above, a similar behaviour of the optical features can be expected when the sphere dielectric constant decreases while the sphere radius remains unchanged. This can be observed in Fig. 7, where the reflectance spectra for close-packed square and triangular lattices are plotted for decreasing values of the dielectric constant of the spheres. Again, a pattern of alternating broad and sharp peaks is obtained, TM and TE in origin, and these peaks move towards higher energies as the sphere dielectric constant decreases.

In conclusion, we have explained the origin of the peaks observed in the reflectance spectra of self-standing monolayers, composed of periodically arrayed spheres of low dielectric constant, for those energies below the onset of diffraction spots. This task has been performed by analysing the contributions of dipolar, quadrupolar and octupolar fields in the spherical wave expansion used in the vector KKR method. We have found that the peak structure in the reflectance spectrum (and dips in the transmittance spectrum) originates from the resonance of TM and TE modes inside the monolayer, not from the Mie resonances of a single sphere, with a broad reflectance peak associated with a TM mode resonance and a sharp reflectance peak associated with a TE mode resonance. These qualitative features are universal because they are independent of the particular geometry of the lattice: they emerge as a consequence of an existing 2D periodicity inside the monolayer and its associated internal electrodynamics, regardless of the geometric details of the lattice.

Acknowledgments

This work has been realized in the framework of a joint Spanish-Argentinian cooperation project CSIC-CONICET (Grant 2005AR0070). The authors are very grateful to Hernán Míguez (Instituto de Ciencia de Materiales de Sevilla) for useful discussions. Support from Consejo Nacional de Investigaciones Científicas y Técnicas (CONICET), Universidad de Buenos Aires (UBA) and Agencia Nacional de Promoción Científica y Tecnológica (ANPCYT-BID 802/OC-AR03-14099) is also acknowledged.

**Figure captions**

FIG. 1. (Color online) (a) Specular reflectance, $R_{0,0}$ (black solid line), and total reflected light, R (red dashed line), spectra of a square lattice of close-packed spheres of dielectric constant $\varepsilon_s$ = 2.4964 (radius $S$ = 0.5$a$) in air. (b) Total reflected diffracted light, $\Sigma_{(p,q)\neq(0,0)} R_{p,q}$, spectrum. The vertical dashed line at $a/\lambda$ = 1 indicates the onset of diffraction.

FIG. 2. (Color online) (a) Specular reflectance, $R_{0,0}$ (black solid line), and total reflected light, R (red dashed line), spectra of a triangular lattice of close-packed spheres of dielectric constant $\varepsilon_s$ = 2.4964 (radius $S$ = 0.5$a$) in air. (b) Total reflected diffracted light, $\Sigma_{(p,q)\neq(0,0)} R_{p,q}$, spectrum. The vertical dashed line at $a/\lambda$ = 1.155 indicates the onset of diffraction.

FIG. 3. (Color online) (a) Reflectance spectra of a square lattice in the range 0.7 ≤ $a/\lambda$ ≤ 1.0 for various values of the sphere radius $S$. (b) Reflectance spectra of a square lattice in the smaller range 0.98 ≤ $a/\lambda$ ≤ 1.0 for various values of the sphere radius $S$. The dielectric constant of the spheres is $\varepsilon_s$ = 2.4964 and the embedding medium is air.

FIG. 4. (Color online) (a) Reflectance spectra of a triangular lattice in the range 0.8 ≤ $a/\lambda$ ≤ 1.155 for various values of the sphere radius $S$. (b) Reflectance spectra of a triangular lattice in the smaller range 1.1 ≤ $a/\lambda$ ≤ 1.155 for various values of the sphere radius $S$. The dielectric constant of the spheres is $\varepsilon_s$ = 2.4964 and the embedding medium is air.

FIG. 5. (Color online) (a) Comparison between the reflectance spectrum ($L_{MAX}$ = 9) and its dipolar contribution ($L_{MAX}$ = 1) of a square lattice for energies just below the onset of diffraction at $a/\lambda$ = 1.0. (b) TM and TE dipolar ($L_{MAX}$ = 1) contributions to the reflectance spectrum in (a) of a square lattice. (c) Comparison between the reflectance spectrum ($L_{MAX}$ = 9) and its dipolar contribution ($L_{MAX}$ = 1) of a triangular lattice for energies just below the onset of diffraction at $a/\lambda$ = 1.155. (d) TM and TE dipolar ($L_{MAX}$ = 1) contributions to the reflectance spectrum in (c) of a triangular lattice. The sphere radius is $S$ = 0.2$a$ and the sphere dielectric constant is $\varepsilon_s$ = 2.4964, the embedding medium is air.

FIG. 6. (Color online) (a) Comparison between the reflectance spectrum ($L_{MAX}$ = 9) and its quadrupolar and octupolar contributions ($L_{MAX}$ = 2 and 3) of a close-packed square lattice for energies below the onset of diffraction at $a/\lambda$ = 1.0. (b) TM and TE octupolar ($L_{MAX}$ = 3) contributions to the reflectance spectrum in (a) of a close-packed square lattice. (c) Comparison between the reflectance spectrum ($L_{MAX}$ = 9) and its quadrupolar and octupolar contributions ($L_{MAX}$ = 2 and 3) of a close-packed triangular lattice for energies below the onset of diffraction at $a/\lambda$ = 1.155. (d) TM and TE octupolar ($L_{MAX}$ = 3) contributions to the reflectance spectrum in (c) of a close-packed triangular lattice. The sphere radius is $S$ = 0.5$a$ and the sphere dielectric constant is $\varepsilon_s$ = 2.4964, the embedding medium is air.

FIG. 7. (Color online) (a) Reflectance spectra of a close-packed square lattice for decreasing values of the sphere dielectric constant $\varepsilon_s$. (b) Reflectance spectra of a close-packed triangular lattice for decreasing values of the sphere dielectric constant $\varepsilon_s$. The sphere radius is $S$ = 0.5$a$ and the embedding medium is air.

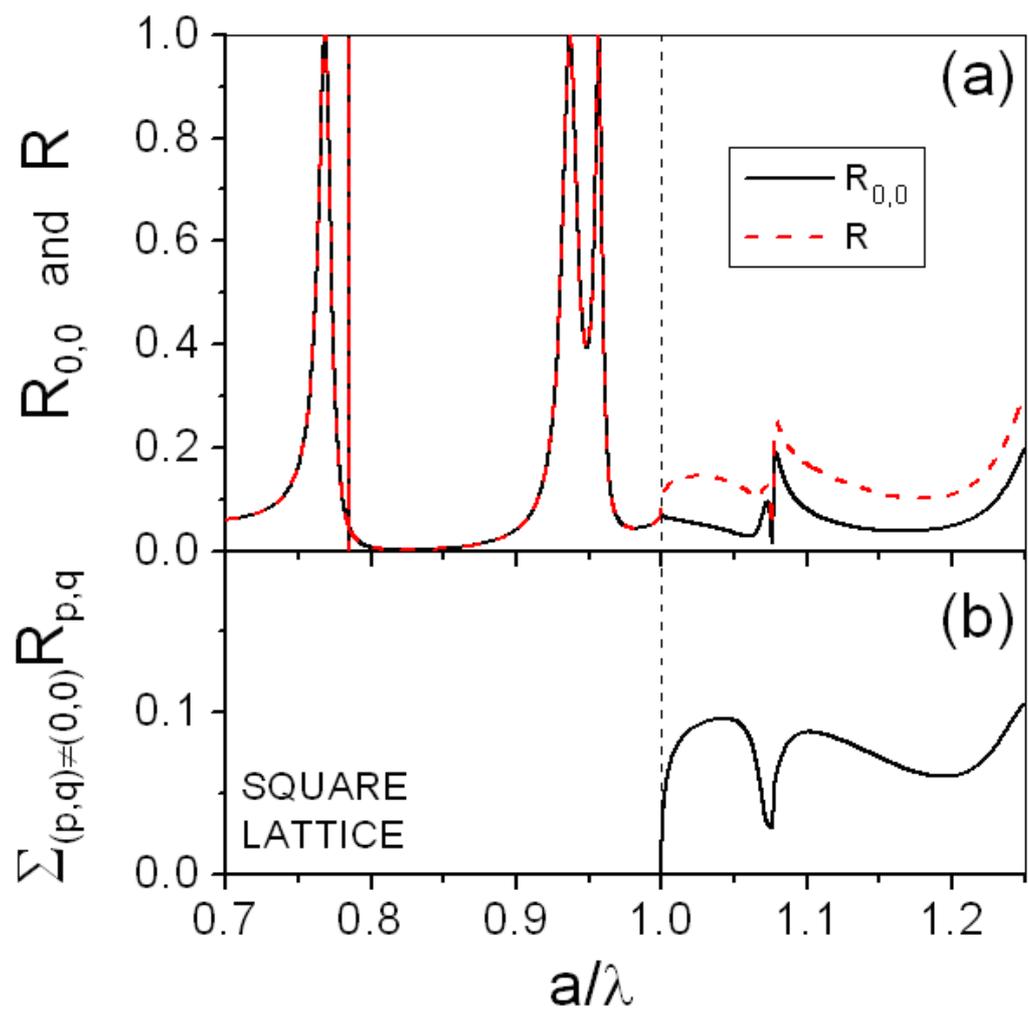

Figure 1. L. Dorado et al.

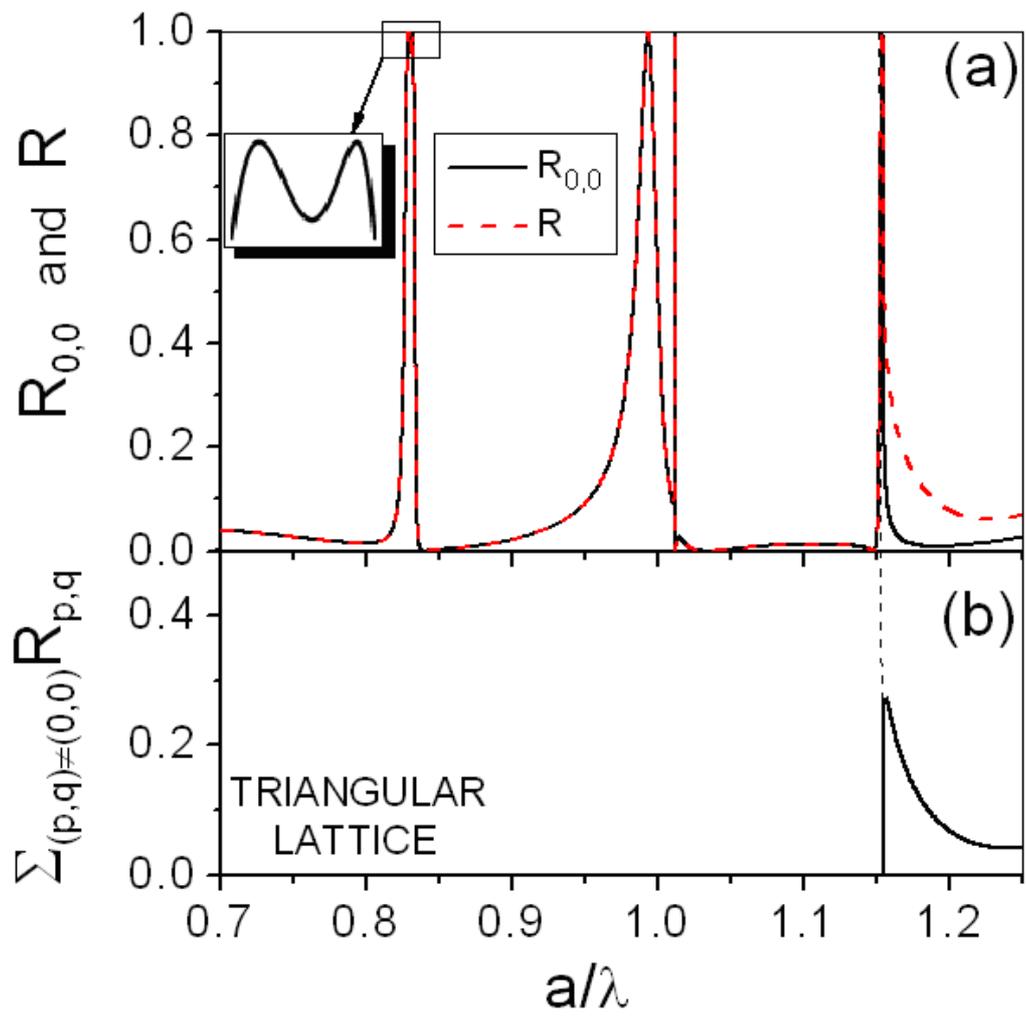

Figure 2. L. Dorado et al.

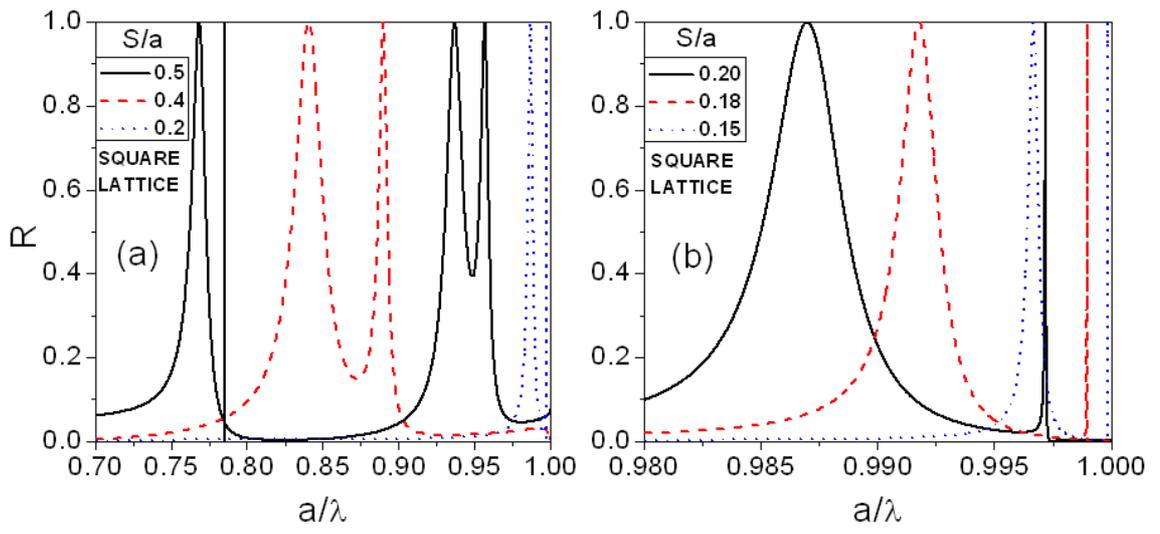

Figure 3. L. Dorado et al.

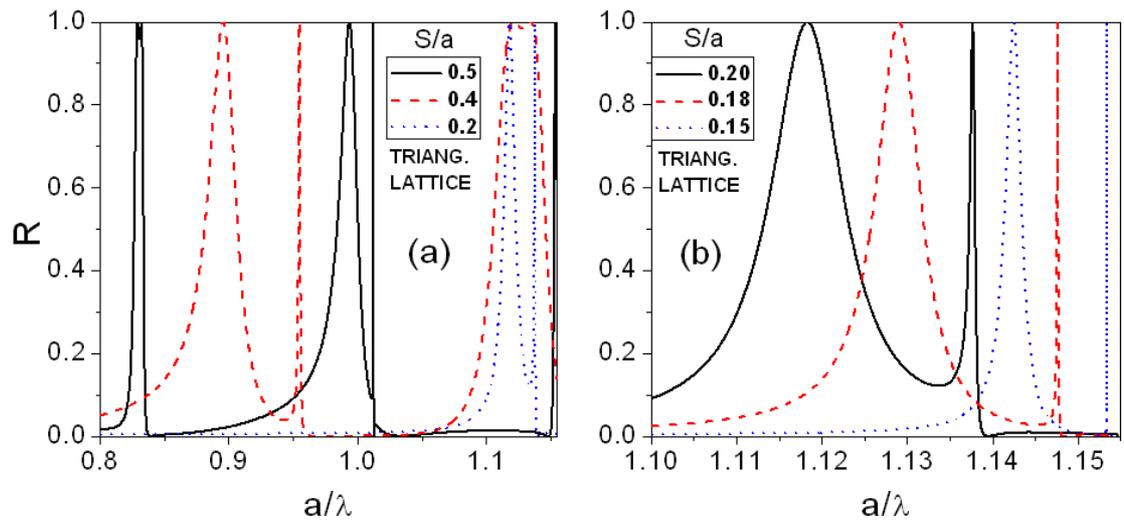

Figure 4. L. Dorado et al.

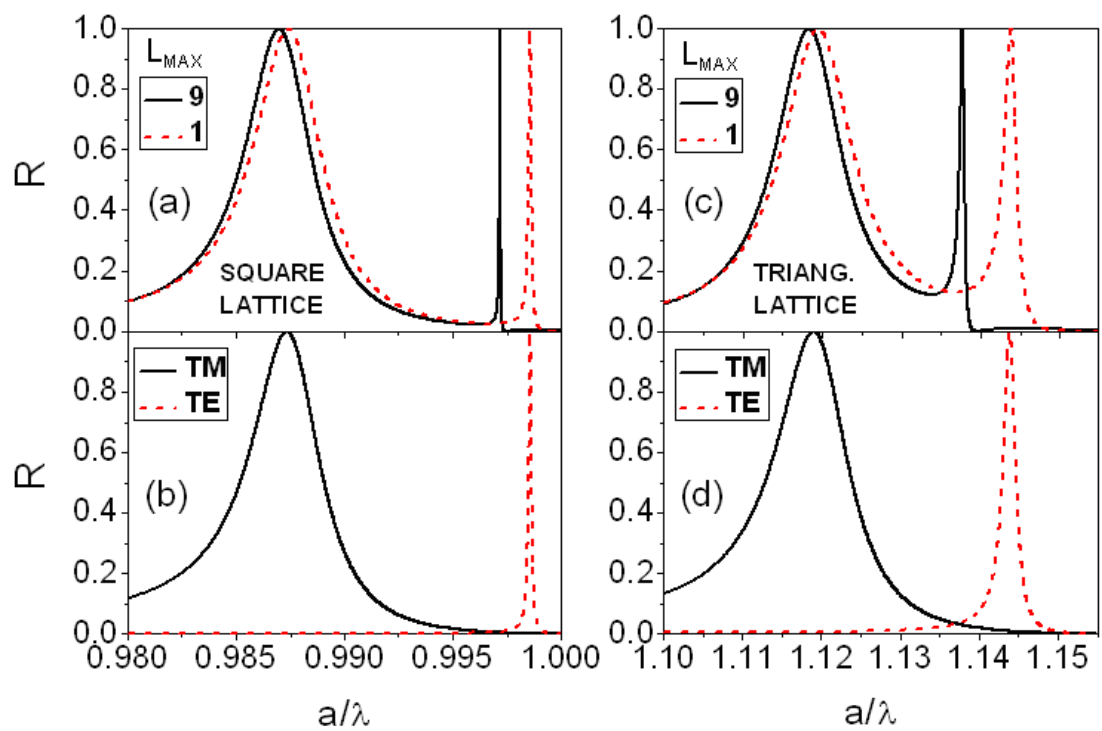

Figure 5. L. Dorado et al.

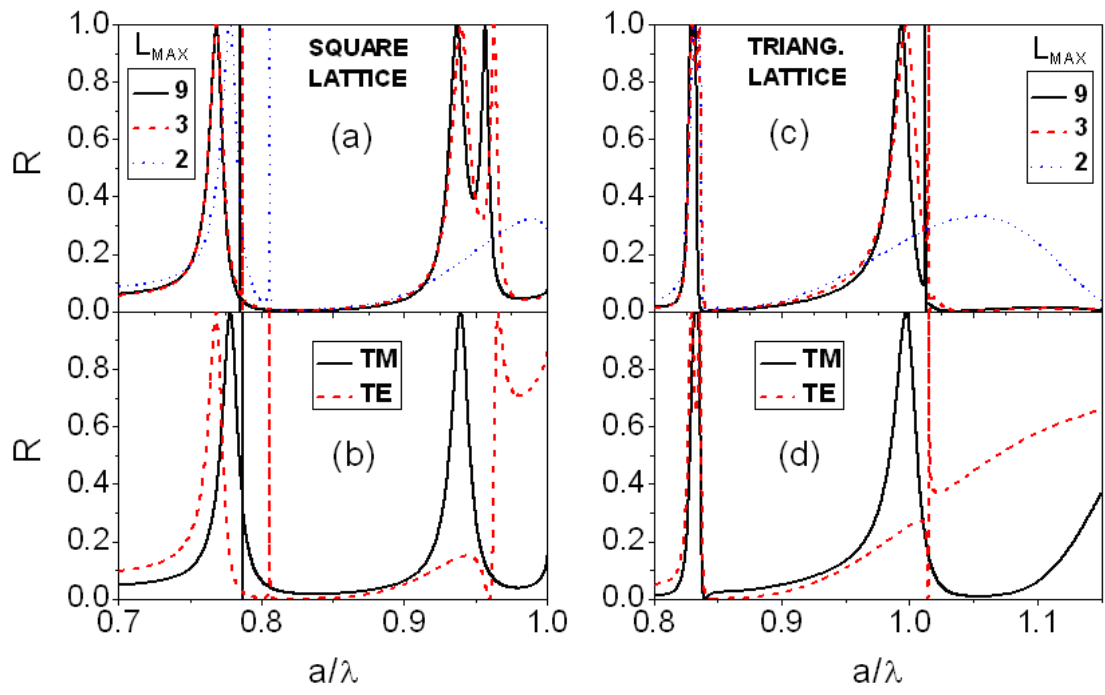

Figure 6. L. Dorado et al.

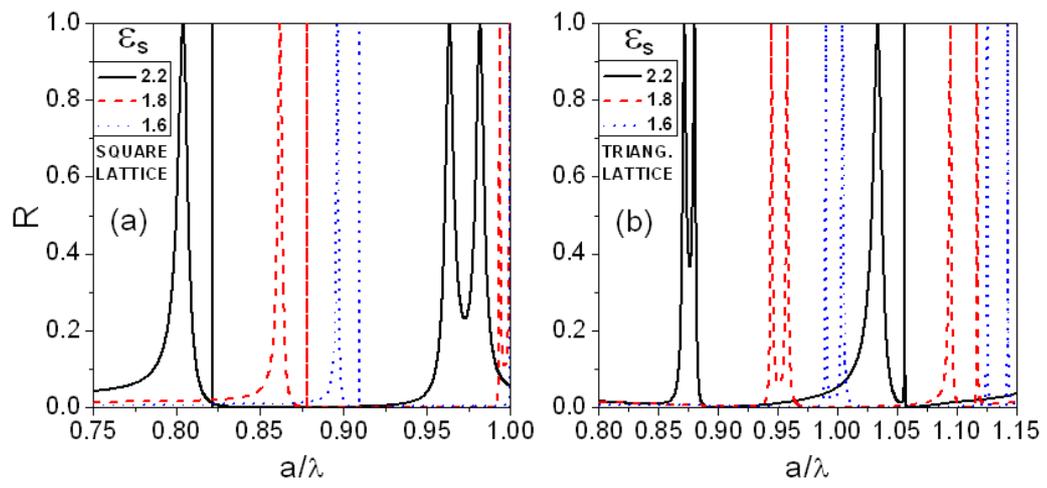

Figure 7. L. Dorado et al.

**References**

[1] M. Inoue, *Phys. Rev. B* **36**, 2852 (1987).

[2] H. Miyazaki, and K. Ohtaka, *Phys. Rev. B* **58**, 6920 (1998).

[3] K. Ohtaka, Y. Suda, S. Nagano, T. Ueta, A. Imada, T. Koda, J. S. Bae, K. Mizuno, S. Yano, and Y. Segawa, *Phys. Rev. B* **61**, 5267 (2000).

[4] H. Míguez, V. Kitaev, G. Ozin, *Appl. Phys. Lett.* **84**, 1239 (2004).

[5] J. F. Galisteo-López, C. López, *Phys. Rev. B* **70**, 035108 (2004).

[6] F. García-Santamaría et al., *Phys. Rev. B* **71**, 195112 (2005).

[7] N. Stefanou, A. Modinos, *J. Phys.: Condens. Matter* **3** 8135 (1991).

[8] T. Kondo et al., *Phys. Rev. B* **66**, 033111 (2002).

[9] T. Kondo et al., *Phys. Rev. B* **70**, 235113 (2004).

[10] Y. Kurokawa, Y. Jimba, H. Miyazaki, *Phys. Rev. B* **70**, 155107 (2004).

[11] M. Born and E. Wolf, Principles of Optics (Pergamon, Oxford, 1965), p. 633.

[12] T. Ochiai, K. Sakoda, *Phys. Rev. B* **64**, 045108 (2001).

[13] H. Miyazaki, Y. Jimba, *Phys. Rev. B* **62**, 7976 (2000).

[14] K. Mukaiyama, K. Takeda, H. Miyazaki, Y. Jimba, and M. Kuwata-Gonokami, *Phys. Rev. Lett.* **82**, 4623 (1999).

[15] J.D. Jackson, *Classical Electrodynamics* (Wiley, New York,1975).

[16] A. Modinos, *Physica A* **141**, 575 (1987).

[17] N. Stefanou, V. Yannopapas, A. Modinos, *Comput. Phys. Commun.* **113**, 49 (1998); **132**, 189 (2000).

[18] K. Ohtaka, *J. Phys. C: Solid State Phys.* **13**, 667 (1980).

[19] N. Stefanou, V. Karathanos, A. Modinos, *J. Phys.: Condens. Matter* **4** 7389 (1992).

[20] J.B. Pendry, *Low Energy Electron Diffraction* (Academic Press, London, 1974).